\documentclass[oldversion]{aa}
\usepackage{epsfig}
\usepackage{natbib}
\usepackage{lscape}
\usepackage{amssymb}
\usepackage{epstopdf}
\usepackage{longtable}
\usepackage{color}
\usepackage{fancyheadings}
\bibpunct{(}{)}{;}{a}{}{,}

\begin{document}
\title{Impact of occultations of stellar active regions on transmission spectra}

\subtitle{Can occultation of a plage mimic the signature of a blue sky?}

\author{ M. Oshagh\inst{1,2} \and N. C. Santos\inst{1,2} \and D. Ehrenreich \inst{3} \and N. Haghighipour\inst{4} \and
 P. Figueira\inst{1} \and A. Santerne\inst{1} \and M. Montalto\inst{1}
}

\institute{
Centro de Astrof{\'\i}sica, Universidade do Porto, Rua das Estrelas, 4150-762 Porto,
Portugal \\
email: {\tt moshagh@astro.up.pt}
\and
Departamento de F{\'i}sica e Astronomia, Faculdade de Ci{\^e}ncias, Universidade do
Porto,Rua do Campo Alegre, 4169-007 Porto, Portugal
\and
Observatoire de Gen\`{e}ve, Universit{\'e} de Gen\`{e}ve, 51 chemin des Maillettes, CH-1290 Sauverny, Switzerland
\and
Institute for Astronomy and NASA Astrobiology Institute, University of Hawaii-Manoa,
2680 Woodlawn Drive, Honolulu, HI 96822,USA
}

\date{Received XXX; accepted XXX}

\abstract {Transmission spectroscopy during planetary transits, which is based on the measurements of the variations 
of planet-to-star radius ratio as a function of wavelength, is a powerful technique to
study the atmospheric properties of transiting planets. One of the main limitation of this technique is the effects of
stellar activity, which up until now, have been taken into account only by assessing the effect of non-occulted stellar spots on the 
estimates of planet-to-star radius ratio. In this paper, we study, for the first time, the impact of the occultation of a
stellar spot and plage on the transmission spectra of transiting exoplanets. We simulated this effect by generating a large number of 
transit light curves for different transiting planets, stellar spectral types, and for different wavelengths. 
Results of our simulations indicate that the anomalies inside the transit light curve can lead to a significant underestimation 
or overestimation of the planet-to-star radius ratio as a function of wavelength. At short wavelengths, the effect can reach
to a difference of up to $10\%$ in the planet-to-star radius ratio, mimicking the signature of light scattering in the 
planetary atmosphere. Atmospheric scattering has been proposed to interpret the increasing slopes of transmission spectra 
toward blue for exoplanets HD 189733b and GJ 3470b. Here we show that these signatures can be alternatively interpreted by 
the occultation of stellar plages. Results also suggest that the best strategy to identify and quantify the effects of stellar 
activities on the transmission spectrum of a planet is to perform several observations during the transit epoch at the same wavelength.
This will allow for identifying the possible variations in transit depth as a function of time due to stellar activity variability.}


\authorrunning{M. Oshagh et al.}
\titlerunning{Impact of occultation of the stellar active region on the transmission spectroscopy}
\maketitle

\section{Introduction}

Most attempts in characterizing exoplanetary atmospheres have been made using the transmission spectroscopy 
(multiband photometry) of transiting exoplanets. In this approach,  observations of planetary transits 
in different wavelengths are used to determine planet-to-star radius ratio, 
${R_{p}}/{R_{\ast}}= k(\lambda)$, as a function of color. 
The inferred wavelength dependence of this quantity is the result of differential absorption in the 
planetary atmosphere \citep{Seager-00, Brown-01, Charbonneau-02} which has been used to constrain 
different atmospheric composition models [e.g., GJ 3470b \citep{Demory-13,Fukui-13, Nascimbeni-13, Crossfield-13}, 
GJ 1214b \citep{Kreidberg-14}, HD 209458b \citep{Desert-08}, and HD 189733b \citep{Pont-07, Ehrenreich-07, Pont-08, 
Sing-09, Sing-11, Pont-13}]. 

Although the magnitude of the planet-to-star radius ratio is strongly affected by the activity of the central 
star and its corresponding surface features \citep{Czesla-09, Oshagh-13a}, except for the systems of GJ436, GJ3470, and HD 189733 
\citep{Pont-07,Knutson-11,Sing-11,Nascimbeni-13,Pont-13}, in most studies using  transmission spectroscopy, the impact 
of stellar features has not been taken into account. Even in the cases of GJ436 and GJ3470, only the impact of non-occulted 
stellar spots has been discussed. The system of  HD 189733 is the only case in which the effect of the occulted stellar spot
inside the transit has been taken into account.

The main objective of this paper is to examine the possible impact of the occultation of stellar activity features
(such as spots and plages) by transiting planets, on transmission spectra. 
In particular, and for the first time, we explore the impact of the occultation
of plages (bright regions in the stellar chromosphere). We perform a large number of simulations to quantify the
impact of this effect on the transit depth measurements in different wavelengths and for various physical 
configurations. To demonstrate the application of our results, we will apply our methodology to the planetary systems
of two active stars HD 189733 and GJ 3470. Due to their low density and large planet-to-star radius ratio, 
these systems are among the most favorable targets for atmospheric characterization purposes. The measurements 
of planet radii in these systems have shown an excess in the short wavelength regime of 300--800 nm. Several 
studies of the entire transmission spectra of these two planets have contributed those observed excesses to Rayleigh
scattering processes in the planets' atmospheres. In this paper, we investigate the possibility of explaining these 
access features by taking the effects of stellar spots and plages occultation into account.

In Sect. 2, we present the details of our models that are used to produce light curves of a planet transiting
plages and spots. In Sect. 3, we apply our models to different configurations of stars with different spectral 
types, stellar activity features, and for planet radii corresponding to wavelengths ranging from $400$ nm to 
$4500$ nm. In Sect. 4 we reanalyze the transmission spectra of HD 189733b as reported by \citet{Pont-13} and 
GJ 3470b as reported by \citet{Nascimbeni-13}, and explore possible scenarios which could reproduce the 
same transmission spectra by invoking only occultation of stellar plages. In Sect. 5, we conclude our 
study by summarizing the results and discussing their implications.

\section{Description of the Model}

We considered a transiting system with a planet in a 3-day orbit. We chose the central star to be 
of spectral types M and G, and took the planet to be of Jupiter (J) and Neptune (N) sizes. The G star was chosen to
be Sun-like and the radius of the M star was taken to be 0.7 solar-radii. Both stars have a rotational period of 9 days. The planet-to-star 
radius ratios in these systems are $0.035$, $0.1$, $0.05$, and $0.15$ for the NG, JG, NM, and JM systems, respectively.
We chose this systems because the large values of their planet-to-star radius ratio make them favorable for 
planet atmosphere studies. 

To study the effects of spots and plages, we assigned them a filling factor defined as

\begin{equation}
f = \frac{A_s}{A_{\ast}}= \left(\frac{R_s}{R_{\ast}}\right)^{2}\,.
\end{equation}

\noindent
In this equation, $A_s$ is the area of the stellar activity feature, $R_s$ is its radius, and $A_{\ast}$ is 
is the area of the stellar disk.
In applying equation (1) to spots, we considered $f$ to be 0.025\% \, {\rm and}\, 1\% with 1\% being the largest 
filling factor of a Sun spot \citep{Solanki-03, Meunier-10}. For plages, we considered 
$f=0.025\%\,,1\%\, {\rm and}\,6.25 \%$, 
in which the maximum value coincides with the maximum filling factor of a Sun's plage \citep{Meunier-10}.
We assumed that at the time of the overlap between the transiting planet and the stellar activity feature, 
the feature is on the star's equator and at a longitude equal to  0.5 stellar radii. As shown by \citet{Oshagh-13a}, 
the maximum underestimation of the planet radius in the transit light curve, with a spot anomaly inside 
the transit, occurs when the spot anomaly appears in the center of the transit light curve.

\subsection{Generating Light Curves}

To generate the light curves of our system, we used the publicly available software SOAP-T \citep{Oshagh-13b}. 
This software has the capability of producing light curves of systems where a transiting planet orbits
a rotating star with activity features in the form of stellar spots or plages. 
The three wavelength-dependent stellar parameters in this software, namely 
the coefficients of quadratic stellar limb darkening ($u_{1}$ and $u_{2}$) and the relative brightness of 
stellar active regions ($b$) enable one to produce light curves for different values of wavelength.
Table 1 shows the values of $u_{1}$ and $u_{2}$ adopted from the catalog by
\citet{Claret-11} for the values of the wavelength used in our simulations. These wavelengths have been chosen to 
cover the entire visible spectrum as well as the Near- and Mid-Infrared ranges.

\begin{table}[h]

\caption{Quadratic limb darkening coefficients for M and G stars in different wavelength \citep{Claret-11}.}
\begin{center}
\begin{tabular}{c c c c c c }
\hline
\\
$\lambda(nm)$& 400 &800 & 1500& 3000&4500\\
\\
\hline
M-dwarf\\
\hline
$u_{1}$ &0.45&0.43&0.40&0.05&0.05\\
$u_{2}$ &0.36&0.34&0.36&0.20&0.18\\
\\
\hline
G-dwarf\\
\hline
$u_{1}$ &0.70&0.30&0.10&0.07&0.05\\
$u_{2}$ &0.18&0.30&0.32&0.14&0.12\\

\hline
\end{tabular}
\end{center}
\label{default}
\end{table}%

To determine the value of $b$, we note that to the first-order of approximation,
this quantity can be written as

\begin{equation}
b(\lambda)=\frac{{\rm exp}[{hc}/{\lambda K_{\rm B} T_{\ast}}]-1}{{\rm exp}
[{hc}/{\lambda K_{\rm B} T_{\rm s}}]-1}\,.
\end{equation}

\noindent
In this equation, $h$ is the Planck constant, $c$ is the speed of light, $K_{\rm B}$ is the Boltzmann constant, $T_{\ast}$
is the stellar temperature, and $T_{\rm s}$ denotes the temperature of the activity feature. We considered
$T_\ast$= 5780 K and 3000 K for our G and M stars, respectively.
For any star with a temperature within this range, the amplitude of the effect of the occultation of 
its active region will be between those of these two stars (Figure 3). 

To obtain the value of $T_s$ for a stellar spot, we used Figure 7 of \citet{Berdyugina-05} where this author 
presented the observed temperature of spots for stars with different spectral types. For plages, this temperature has only 
been measured for the Sun \citep{Worden-98, Unruh-99, Meunier-10} which showed the maximum temperature contrast of 300 K for a plage at the Sun's limb and  the minimum of 100 K at the center of the Sun. We note that the spots on the surface of the Sun show a limb darkening effect. That is, a Sun spot at the 
center of the Sun's disk shows higher contrast compared to the one on the Sun's limb 
\citep{Unruh-99, Meunier-10}. On the contrary, a Sun plage shows a limb brightening behavior, 
which means that a Sun plage on the Sun's limb displays higher contrast compared to that on the center 
of the Sun's disk \citep{Meunier-10}. The values for the temperature of spots and plages in this study have been 
chosen by considering these opposite behavior of the spot's limb darkening and the plage's limb brightening. 
This enabled us to take into account the impact of the position of a stellar spot/plage on its maximum 
temperature-contrast when simulating our systems. For instance, a stellar spot in the center of the 
stellar disk could reach to the maximum possible temperature-contrast with the stellar photosphere whereas
a plage in that location would reach to only a third of the maximum possible plage temperature-contrast. By taking into account that in our simulations we consider that the plages are located at the center of the 
stellar disk therefore we use 100 K temperature contrast for a plage on the surface of our G star, and assumed that on an M star, a plage at its maximum temperature contrast will have
similar temperature-difference.
Table 2 shows the values of $T_s$ for all our models. The variations of $b(\lambda)$ with wavelength is shown
in Figure 1.

\begin{table}[]

\caption{Detailed parameters of our models.}
\begin{center}
\begin{tabular}{c c c c c c c c}

\hline
\\
Model Number & $T_\ast$& $T_s$   & $f$ & ${R_{p}}/{R_\ast}$  \\
\\
\hline
G dwarf+ spot\\
\hline
1 & 5780 & 4000& 0.25\% & 0.035 \\
2 & 5780 & 4000& 0.25\% & 0.1 \\
3 & 5780 & 4000& 1\% & 0.035 \\
4 & 5780 & 4000& 1\% & 0.1 \\
\hline
G dwarf+ plage\\
\hline
5 & 5780 & 5880& 0.25\% & 0.035 \\
6 & 5780 & 5880& 0.25\% & 0.1 \\
7 & 5780 & 5880& 1\% & 0.035 \\
8 & 5780 & 5880& 1\% & 0.1 \\
9 & 5780 & 5880& 6.25\% & 0.035 \\
10 & 5780 & 5880& 6.25\% & 0.1 \\
\hline
M dwarf + spot\\
\hline
11 & 3000 & 2500& 0.25\% & 0.05 \\
12 & 3000 & 2500& 0.25\% & 0.15 \\
13 & 3000 & 2500& 1\% & 0.05 \\
14 & 3000 & 2500& 1\% & 0.15 \\
\hline
M dwarf + plage\\
\hline
15 & 3000 & 3100& 0.25\% & 0.05 \\
16 & 3000 & 3100& 0.25\% & 0.15 \\
17 & 3000 & 3100& 1\% & 0.05 \\
18 & 3000 & 3100& 1\% & 0.15 \\
19 & 3000 & 3100& 6.25\% & 0.05 \\
20 & 3000 & 3100& 6.25\% & 0.15 \\

\hline
\end{tabular}
\end{center}
\label{default}
\end{table}%

\begin{figure}
    \includegraphics[width=0.45\textwidth]{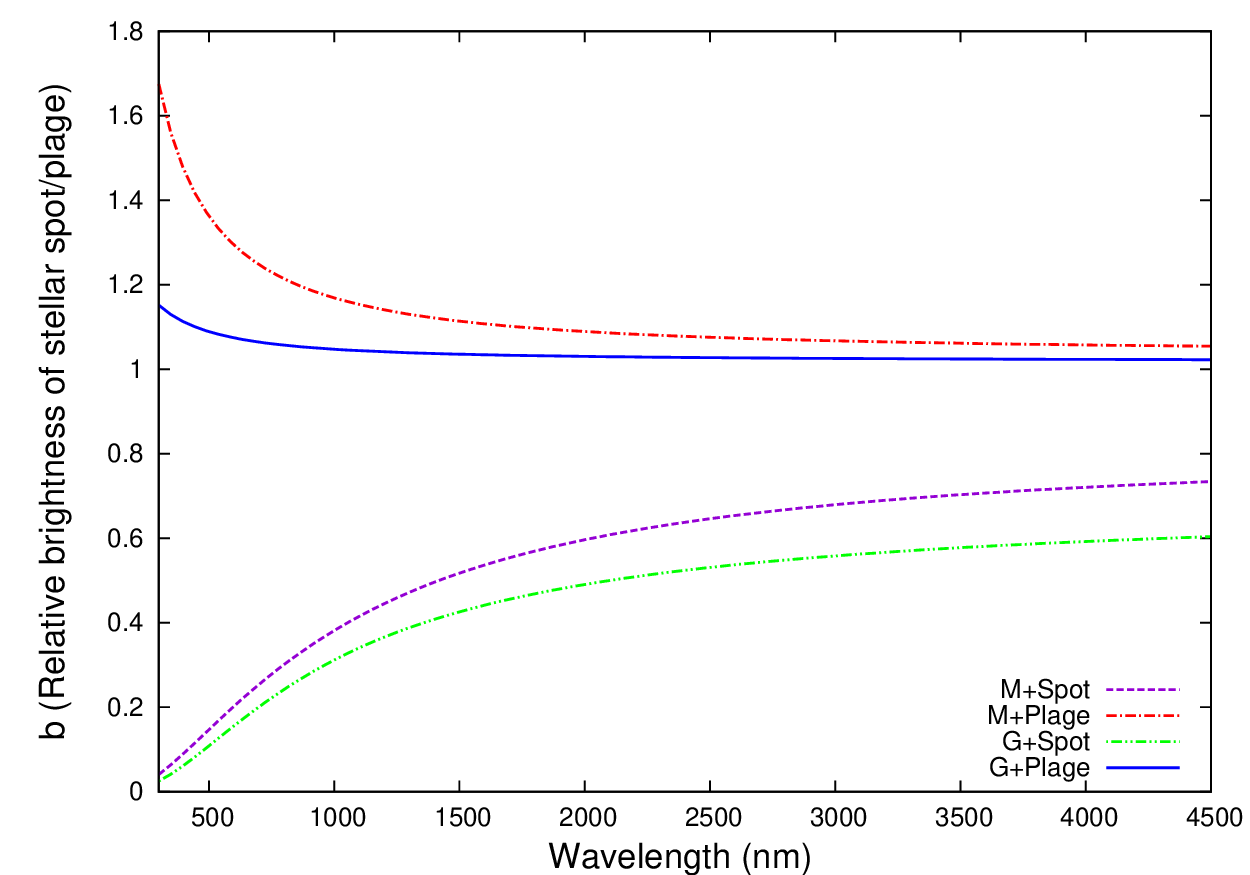}
    \caption{The relative brightness of stellar spots and plages on M and G stars, 
as a function of wavelength. The details of the parameters of spots, plage, and stars are listed in Table 2.}
\end{figure}

\section{Results of Simulations}

We generated a large number of mock transit light curves for our G and M stars considering transiting planets with
radii equal to those of Neptune and Jupiter. As we are interested in determining the impact of the occultation of 
stellar active regions on the estimates of the radius of the transiting planet,
we considered our stars to contain plages or spots, and generated light curves for different values of 
wavelength (see Table 1). As a result, all the generated mock light curves exhibited anomalies in their transit parts.
To examine the effects of these anomalies on the estimates of the planet radius, we fit all these mock light curves
with transit light curve models corresponding to systems in which the planet does not occult the stellar active region.
In each fitted model, we set all the parameters equal to those used to generate the system's initial mock light curve
except for the planet's radius which we allowed to vary freely.
In general, the latter curves show the exact same behavior as those with anomalies
except for inside the transit where the spot/plage occulted anomalies exist. Figure 2 shows a sample of our fitting results
(model 16 in Table 2). The light curve in red corresponds to the system in which the planet occulted a stellar plage. 
The green curve represents the best fitted transit without the occultation of that plage.

It is important to mention that during the fitting procedure, we allow the depth 
of the transit to vary as a free parameter while holding other parameters of the system constant. In that case, for
each light curve with inside transit anomalies, the best fit anomaly-free light curve is used to determine the best planet-to-star
radius ratio. It should be noted that in systems where the planet radius is smaller than the radius of the stellar 
active region (for instance, model 3 in Table 2), only part of the spot/plage would be covered by the planet and as a result, the 
remaining fraction of the spot/plage will affect the out-of-transit light curve similar to an un-occulted spot/plage. However, 
as explained above, in our simulations the only difference between the light curve containing an anomaly and the anomaly-free light curve is only inside their
transits, this out-of-transit light will not affect the results of our fitting.

\begin{figure}[h]
    \includegraphics[width=0.49\textwidth, height=60 mm]{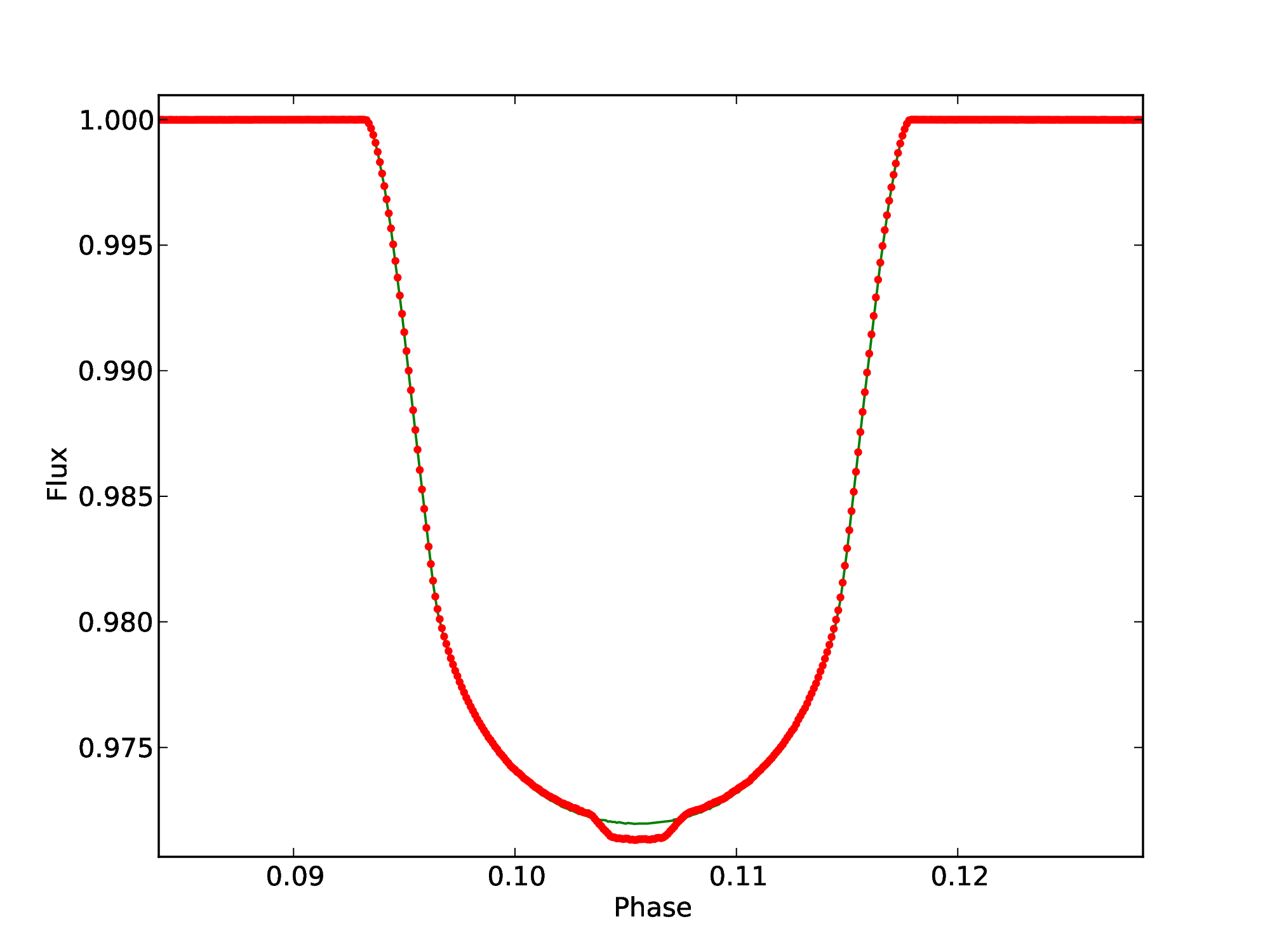}
    \includegraphics[width=0.49\textwidth, height=60 mm]{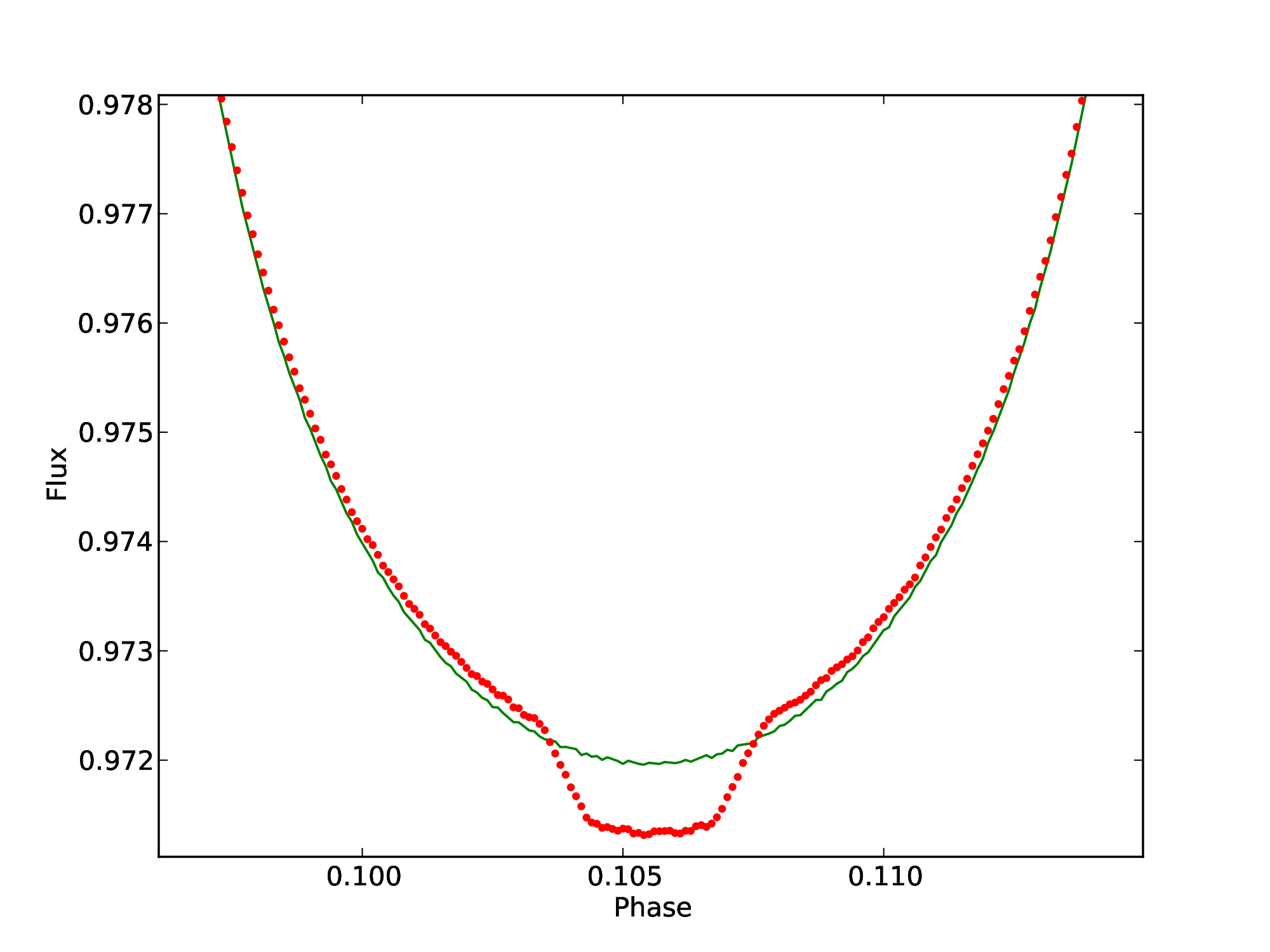}
    \caption{Top: A sample light-curve of our simulations. The system consists of an M star with a 9-day rotational
period with a Jupiter-sized planet $({R_p}/{R_\ast}=0.15)$ in a 3-day orbit. We considered a plage with a filling factor 
of 0.25\% on the surface of the star. The red curve corresponds to the case when the transiting planet occults the plage. 
The green curve shows the best fit transit light-curve without considering the occultation. 
Bottom: Zoomed-in at the bottom of light-curve.}
\end{figure}

Figure 3 shows the relative error in the estimate of the planetary radius obtained from the anomaly-free fitting procedure
compared to the radius of the planet used in generating the mock light curves (i.e., Neptune or Jupiter radii).
As shown here, the underestimation or overestimation of the planet-to-star radius ratio can be quite high 
($ \sim 10\%$), particularly on the blue side of the spectrum where the stellar active regions show higher contrast.
These results also suggest that the stellar spot/plage filling factor seems to play an important role in the estimate of the
planet's radius. For instance, in the models where a Jupiter- or Neptune-sized planet orbit a G-dwarf, the maximum  
effect appears for the largest filling factors corresponding to models 3, 4, 9 and 10 (see Table 2). It should be noted 
that the temperature contrast between a stellar active region and its surrounding area, and the value of the filling 
factor impose a strong degeneracy to the estimate of the planet's radius. That is, a variation in the temperature
contrast of a spot or plage can be compensated by a properly chosen value of the filling factor such that different
combinations of temperature contras/filling factor produce the same value for the planet's radius.

As we noted earlier, in all our simulations, we assumed that the stellar rotation axis is parallel to the plane
of sky, the orbit of the transiting planet is edge-on, and the occultation of a spot/plage occurs when this feature
is on the center of the star's disk. As a result of these assumptions, the values presented here are upper limits.

\begin{figure}
    \includegraphics[width=0.45\textwidth]{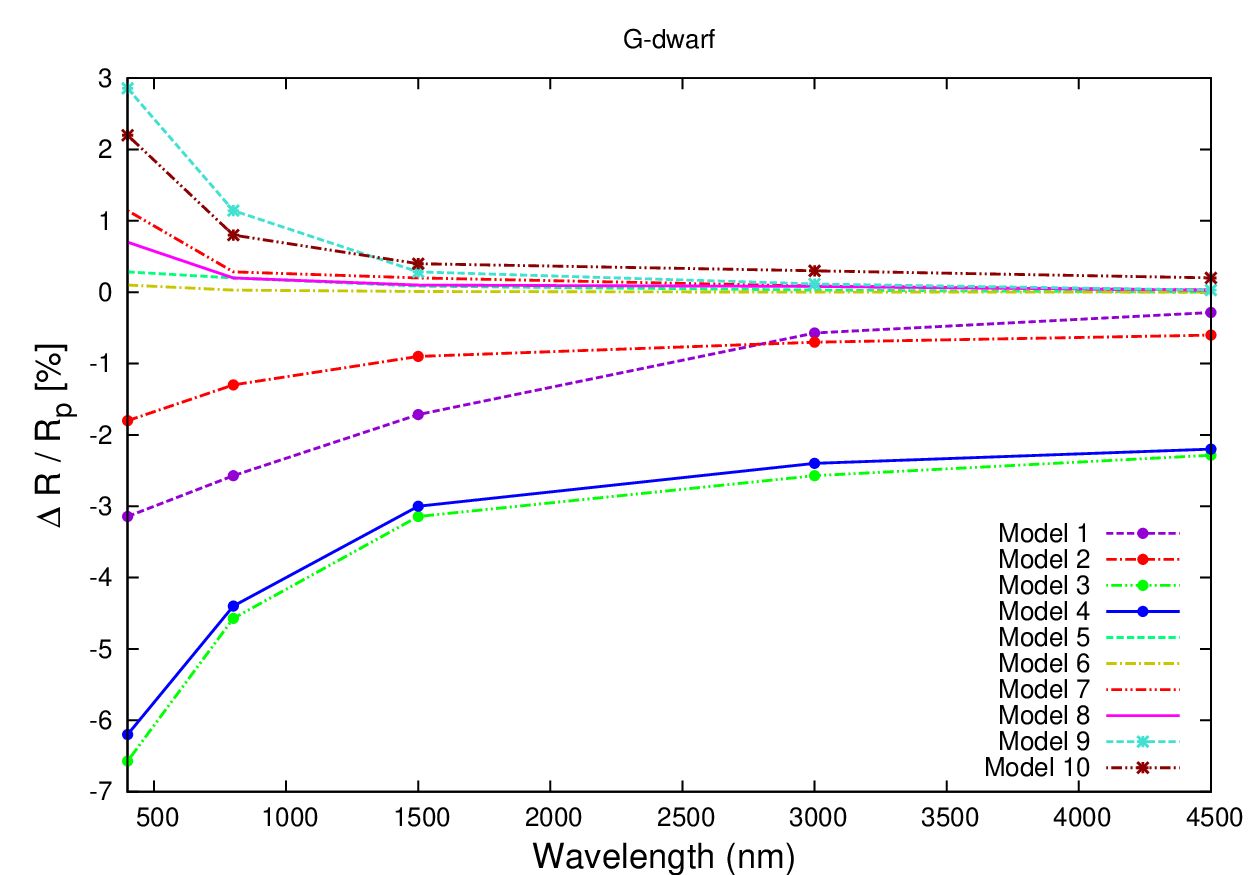}
    \includegraphics[width=0.45\textwidth]{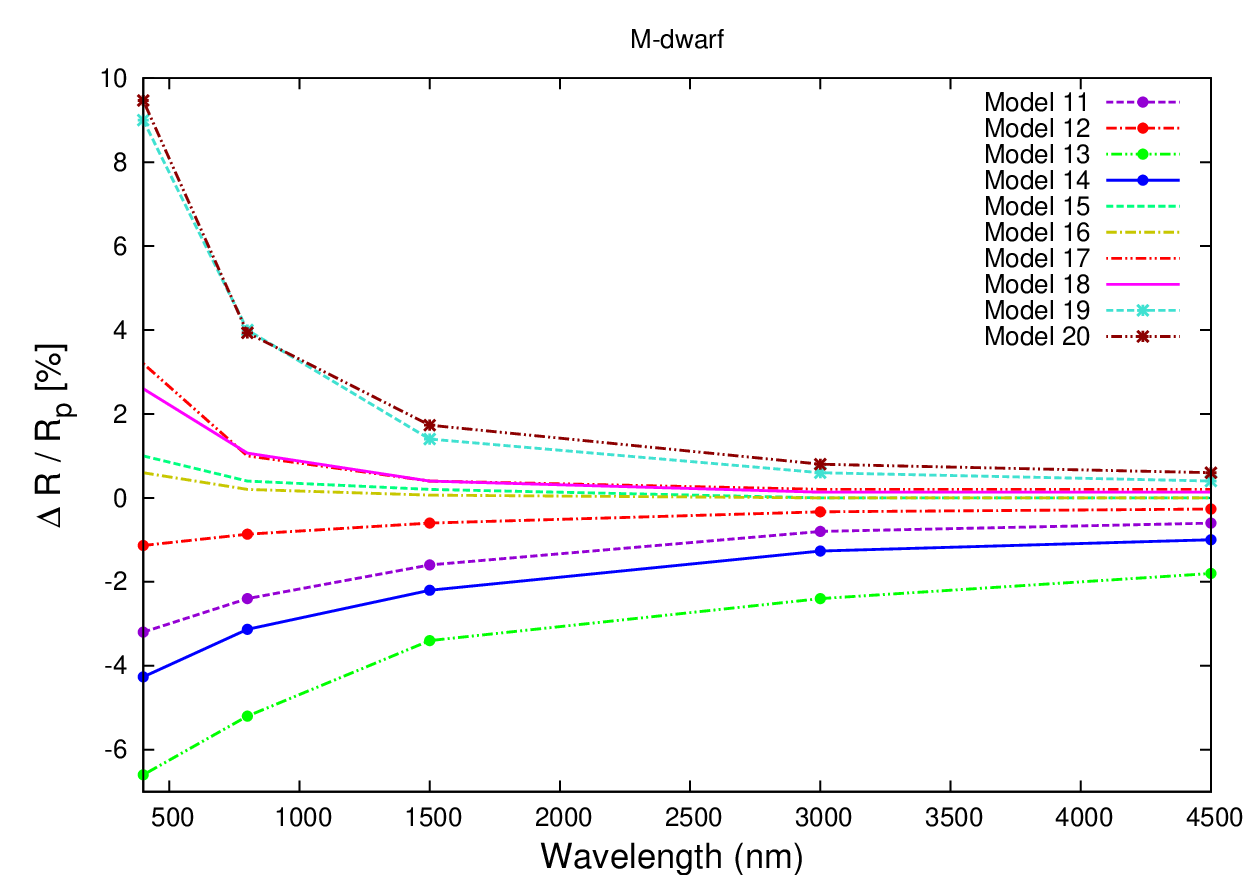} 
    \caption{Graphs of the relative error in the estimate of the radius of a transiting planet as a function of 
wavelength calculated by comparing the value of the planet radius obtained from the best fitted anomaly-free light-curve 
with that obtained from the light-curve taking into account the effects of stellar 
spots and plages occultation. Different colors correspond to different models as described in Table 2.}
\end{figure}

\section{Reanalyzing HD 189733b and GJ 3470b: Do they have blue skies?}

\subsection{HD 189733b}

HD 189733 is a K-dwarf with an effective temperature of 5050 K, surface gravity of $\log g=4.53$, and
brightness of $V\simeq7.7$ \citep{Sing-11}. The short-period planet of this star, HD 189733b, is a 
Jupiter-like body in  a 2.2-day orbit, with a scaled semimajor axis of $a/R_{\ast}=$ 8.92 \citep{Pont-07}.
The ratio of the radius of HD 189733b to that of its host star is 0.155.
The large atmospheric scale height of HD 189733b has made this planet one of the best studied systems 
in the studies of exoplanetary atmospheres \citep[e.g.,][]{Lecavelier-08,Pont-08,Sing-11,Pont-13}. As reported by
\citet{Sing-11} and \citet{Pont-13}, HD 189733b shows an excess in its radius in the entire visual band 
($300-800$ nm), which can be attributed to Rayleigh scattering in the planet's atmosphere. 

HD 189733 is a highly active star. It has a strong X-ray emission and intense chromospheric 
Ca II H and K lines which suggest high level of activity on the surface of this star \citep{Knutson-10, Poppenhaeger-13}. 
During its 12 days stellar rotation, HD 189733 also shows photometric modulation ($\rm Photo_{\rm var}$) of up to 
$\simeq 2\%$ in the mean b and y Str\"{o}mgren bands (510 nm) \citep{Boisse-09, Sing-11}. \citet{Sing-11} have shown that
the light curve of HD 189733 carries clear signatures of stellar spot occultation inside its transit. These authors, and
more recently \citet{Pont-13} studied the effects of occulted and non-occulted stellar spots on the estimates of the 
radius of HD 189733b and found that these anomalies have significant effects on the accurate measurements of the planet's
radius.

We examined the possible influences of stellar plages on the spectra of HD 189733b 
and the measurement of its radius. Following the procedure described in section 3,
we generated mock light curves of HD 189733 for different values of wavelength
considering an overlap between a plage and the transiting planet. We then fitted these mock light curves with 
models in which the planet/plage occultation was not taken into account. Results of our analysis
indicated that the observed transmission spectrum of HD 189733b can be reproduced by considering the planet
occulting a stellar plage with a filling factor of $1.96\%$ and a temperature-contrast of $100$ K. 
As shown in Figure 4, the observed excess in the planet radius in the bluer part 
of spectrum is also well matched by the predictions of the plage-occultation scenario.

By taking into account the above filling factor and the relative brightness of a plage as obtained from equation (2), 
we found that the amplitude of the photometric modulations of  HD 189733 is approximately $0.4\%$ at 510 nm wavelength, which is 
noticeably smaller than the $\sim 2\%$ value obtained from observation. This suggest that our proposed scenario 
(i.e., the occultation of a plage on the surface of the star) is physically viable. We note here that as explained before, 
determining the plage's temperature and filling factor introduce a strong degeneracy in the modeling process.
For instance, if we considered a temperature ratio of $\sim0.87$ for HD 189733 compared to the Sun, and used this ratio 
to estimate the temperature contrast of a plage on the surface of HD 189733 ($\sim 65$ K), the required plage's filling 
factor for reproducing the transmission spectra of HD 189733b would be around  $2.89\%$.

\begin{figure}
    \includegraphics[width=0.49\textwidth, height=58 mm]{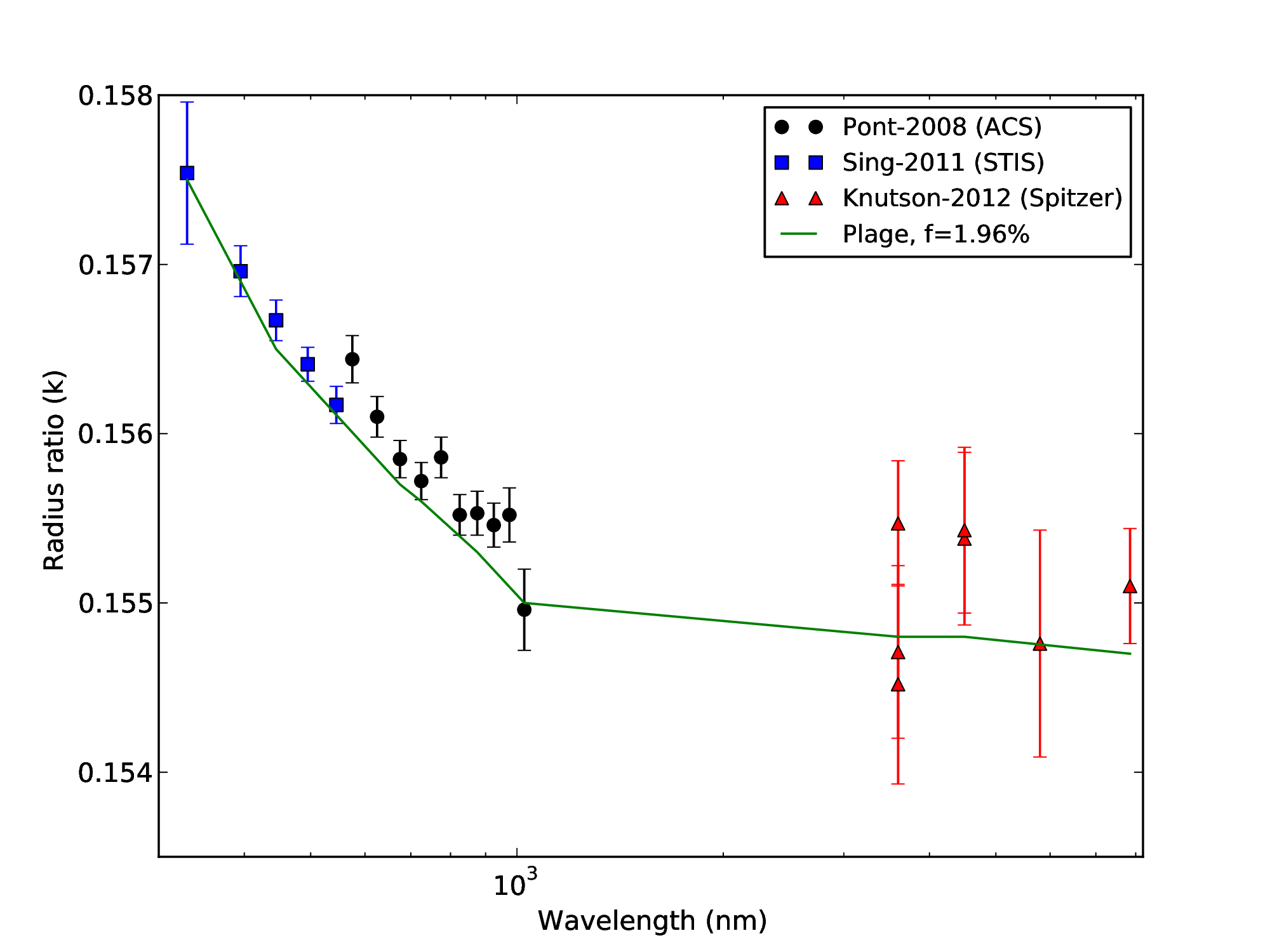}
    \caption{The graph of the observed transmission spectrum of  HD 189733b and the reconstructed transmission
spectrum by assuming the planet overlapping 
a stellar plage. The excess in the estimate of the planet radius-ratio in the blue side of spectrum can 
be reproduced by presuming a plage on the surface of HD 189733. See text for more details.}
\end{figure}

\subsection{GJ 3470b}

GJ 3470 is an M-star with a temperature of 3600 K and surface gravity of $\log g=4.658$ \citep{Demory-13}.
This star is host to a Uranus-mass planet (GJ 3470b) in a 3.34-day orbit with a scaled semimajor axis of
$a/R_{\ast}=$ 14.02 \citep{Fukui-13} and a planet-to-star radius ratio of 0.078. 
GJ 3470b was first discovered with the radial velocity technique using
the HARPS spectrograph. The transits of this planet were later detected through ground-based follow-up observation 
by \citet{Bonfils-12}. Recently, several attempts were made to characterize the atmosphere of GJ 3470b 
using ground- and space-based facilities \citep{Fukui-13,Demory-13,Crossfield-13,Nascimbeni-13,Ehrenreich-14}. 
These observations indicated that the planet's atmosphere has a flat infrared spectrum between 1 and 5 $\mu$m 
suggesting an increase in slope towards the blue side of spectrum ($\lambda\sim 360 $nm), which was interpreted by 
\citet{Nascimbeni-13} to be the result of Rayleigh scattering  in the atmosphere. 
 
As suggested by \citet{Bonfils-12}, because of its slow rotation, GJ 3470 may not be a very active star.  
\citet{Fukui-13} confirmed this finding using photometric observations, and stated that although 
during the course of their 60-day observations, GJ 3470 showed photometric variability of approximately  
$\rm Phot_{\rm var}\simeq 1\%$ at Ic-band (786.5 nm), this star may not still be very active, and those
variabilities can be explained by assuming that GJ 3470 harbors a spot with a filling factor of $1\%$. The latter
motivated \citet{Nascimbeni-13} to study the possible contribution of a non-occulted stellar spot on the spectrum of
GJ 3470b, and they concluded  that their results are not affected by that effect. In this section, we examine whether
an occulted stellar plage can affect the results.

We generated synthetic transit light curves of GJ 3470b considering plage occultation anomalies inside the 
transit, for different values of wavelength. We found that the observed excess in the planet's radius 
in short wavelengths can be explained by the overlap of this planet with a plage with a filling factor of
$2.56 \%$  and temperature contrast of 100 K (Figure 5). The size and relative brightness of such a plage
results in photometric variations of about 1.3\% at 786 nm wavelength, which is compatible with the observed value 1\%.

In larger wavelengths, however, our plage occultation scenario was not able to properly model the results
reported by \citet{Nascimbeni-13} (at the wavelength of 963nm) and the values reported by \citet{Ehrenreich-14} 
for wavelengths between  1170 and 1650 nm. One explanation is that during two successive transits of the planet
at 4520 nm detected by \citet{Demory-13}, the obtained planet radius was systemically offset toward a large value 
causing no significant decrements to be observed in the planet radius in the range of $963-1650$ nm. To examine
this possibility, we assumed that the actual planet-to-star radius ratio is smaller than the reported value (0.077)
and obtained that in this case, the minimum required plage's filling factor will be $\sim 3.61 \%$ with a temperature 
contrast of $100$ K (black solid line in Figure 5).

Note that these values for the plage's temperature and filling factor strongly depend on the choice of the
temperature contrast. For instance, if we use the temperature ratio of GJ 3470 to that of the Sun ($\sim0.62$) to 
estimate the temperature contrast of a plage on the surface of GJ 3470 ($\sim$ 45 K), the plage's filling factors
of  $2.56 \%$  and  $3.61 \%$  will increase to  $7.29 \%$ and $9 \%$, respectively.

\begin{figure}
    \includegraphics[width=0.49\textwidth, height=58 mm]{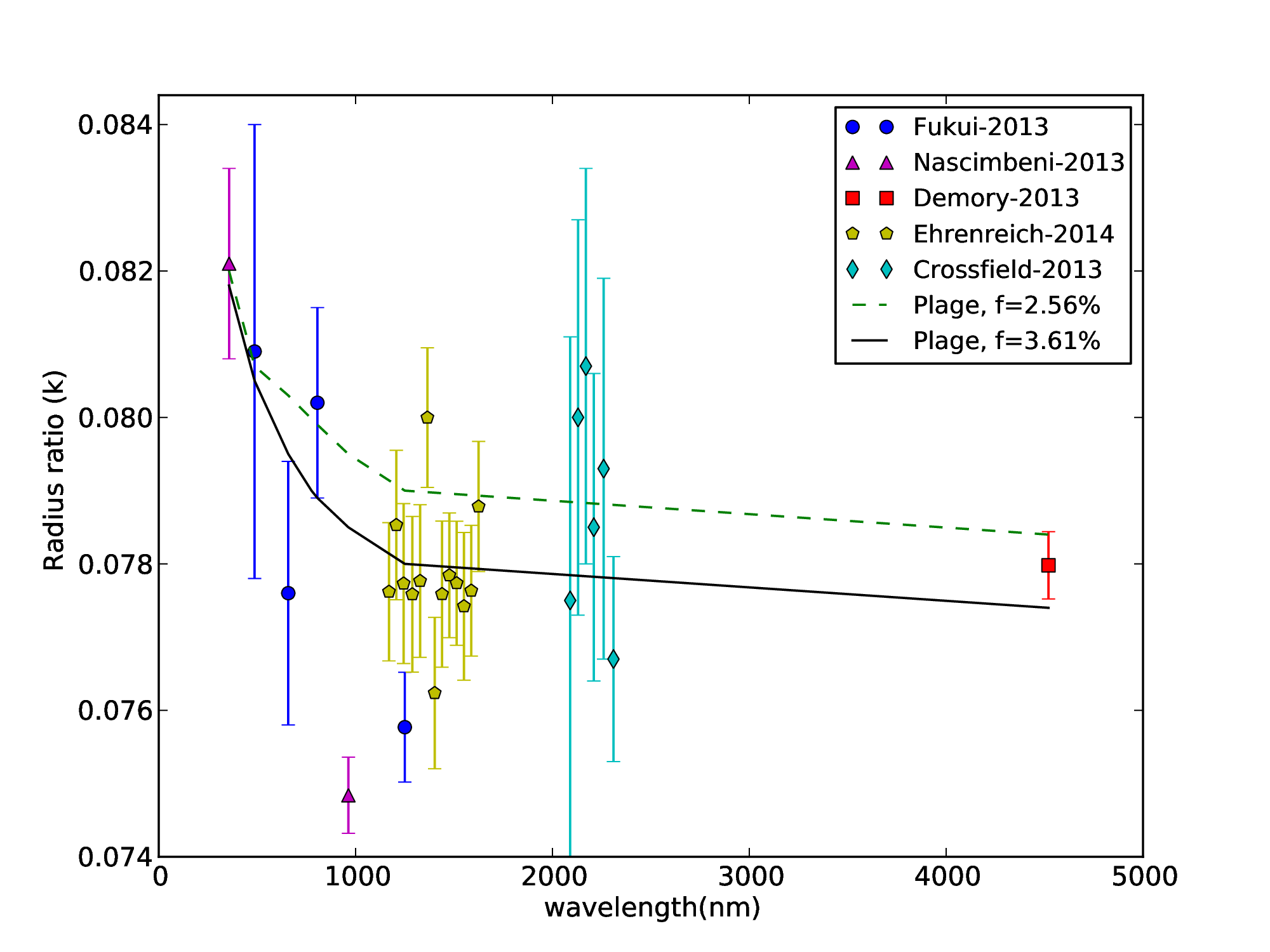}
    \caption{The graph of the observed transmission spectrum of  GJ 3470b and the reconstructed transmission
spectrum by assuming the planet overlapping 
a stellar plage. The excess in the estimate of the planet radius-ratio in the blue side of spectrum can 
be reproduced by presuming a plage on the surface of GJ 3470. See text for more details. }
\end{figure}

\section{Discussion and conclusions}

We present, for the first time, the results of a study of the effect of the occultation of a stellar plages and spots
on the transmission spectroscopy measurements of a planet.
We carried out simulations considering transiting systems with G or M stars, Jupiter- or Neptune-sized planets,
and for different values of a spot/plage's filling factor in different wavelength. Results indicated that there could be 
significant underestimation or overestimation of planet-to-star radius ratio as a function of the wavelength.
The maximum overestimation of planet radius ($10\%$) may occur for the 
occultation of a plage by a planet transiting an M-dwarf in the short wavelength regime. 
Application of our calculations to the systems of the stars HD 189733 and GJ 3470 indicated that the
transmission spectroscopy measurements of the planets of these stars,
and especially the reported excess in their planet-to-star radius 
ratio in the bluer part of the spectra, which were interpreted as the signature of blue sky, can almost exactly be reproduced by assuming the occultation of 
a plage on the surface of these stars.

The results of our study strongly suggest that prior to interpreting the values obtained for the radius of a transiting 
planet in different wavelengths and attempting to set constrains on the planet's atmospheric models, it is crucially important 
to rule out the possible contamination in the measurements due to effects of stellar activities (both occulted and 
non-occulted active regions). The best strategy for doing so is to carry out several observations of a 
transit in a given wavelength and use the variations of transit depth as a function of time 
to assess the impact of potential stellar spot/plage occultation. 
In case of a very active star, which always harbors stellar spots or plages, all transits could be 
affected by the occultation. In that case, the alternative strategy would be to carry out simultaneous multiband 
photometric observations for several hours, which can be corresponded to the half of the stellar rotation period, before and after the transit. These observations will allow to diagnose 
the presence of stellar spots/plages during the transit and can also provide information about the parameters of 
these regions such as their sizes and temperature contrasts. Figure 6 shows an example of the distinct behavior of 
an out-of-transit light curve in different wavelengths in the present of a stellar spot. The system depicted here 
consists of a Jupiter-sized planet with $R_{p}/R_{\ast}=0.15$, orbiting in a 3-day orbit around an 
M-dwarf. The M-dwarf has a rotation period of 9 days, and harbors a spot with 1\% filling factor.

\begin{figure}
    \includegraphics[width=0.45\textwidth]{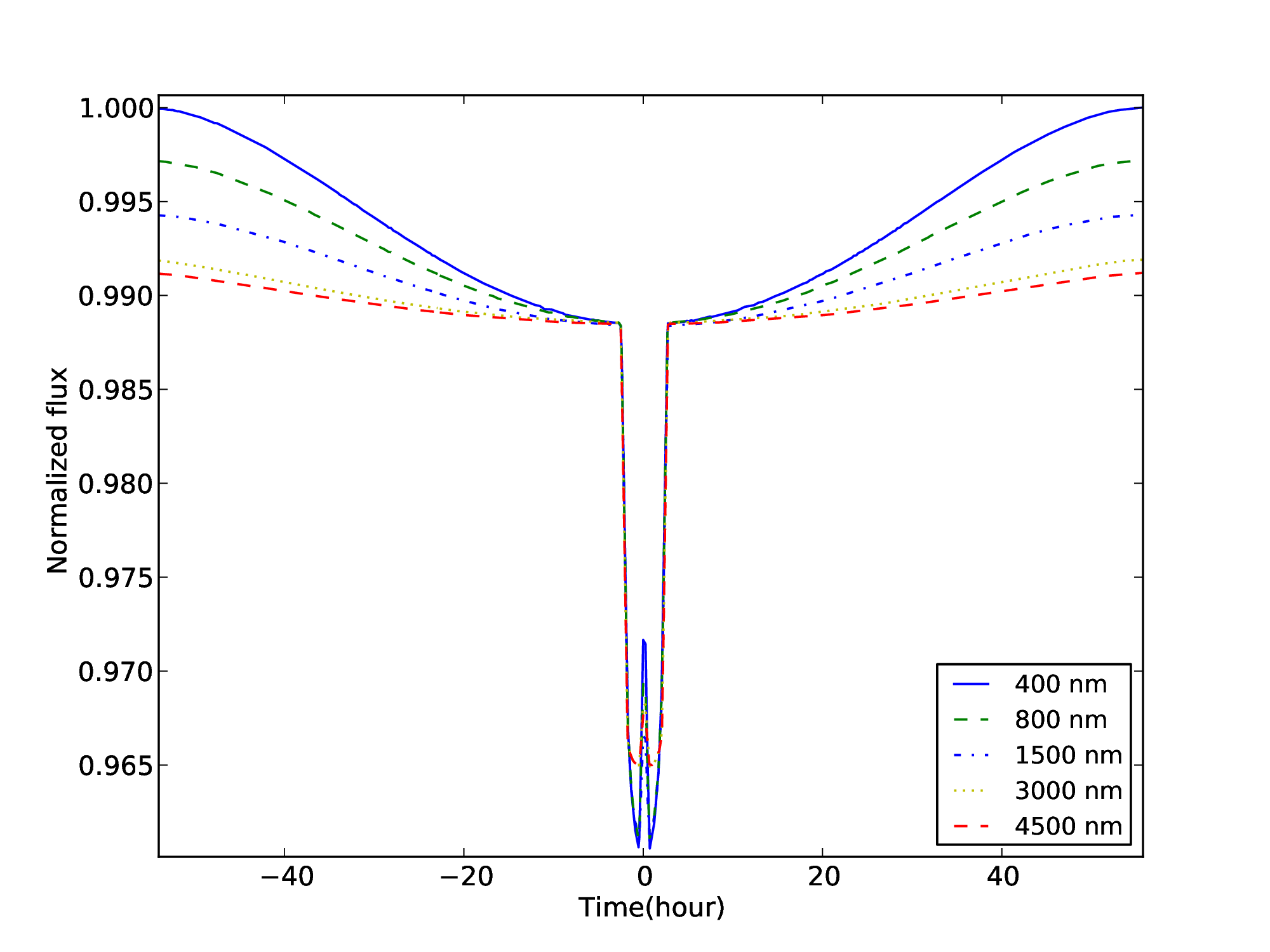}
    \caption{Graphs of out-of-transit light curves for different wavelengths in a system with a stellar spot.}
\end{figure}

We note that the results presented here are based on simple assumptions. For instance, we always 
assumed that the transiting planet would occult the same plage at each epoch. This could be true 
in the case of very active star, such as HD 189733. Our simulations were done by assuming circular spot and plage which might affect the determination of spot/plage's filling factor. We also considered a constant temperature contrast of 
100 K between a plage and its surrounding region on the star's photosphere. However, this value could be
different for different stars. Larger or smaller temperature contrasts will result in lower or higher values
for the filling factor. Finally, we assumed that the plage occultation anomalies appear in the middle of 
transit light-curve so that it would have the maximum impact on planet radius estimations. However,
if these anomalies appear in other locations in the transit light-curve, different values of the filling factor
and/or plage-photosphere temperature contrast will be required to account for the observed transmission 
spectroscopic properties of the system.

\begin{acknowledgements}

We would like to thank the anonymous referee for insightful comments that led to improvements in this paper. 
We acknowledge the support from the European Research Council/European Community under the
FP7 through Starting Grant agreement number 239953, and by Funda\c{c}\~ao para a Ci\^encia e a Tecnologia (FCT) in
the form of grants reference SFRH/BD/51981/2012. NCS also acknowledges
the support from FCT through program Ci\^encia\,2007 funded by FCT/MCTES (Portugal) and POPH/FSE (EC).  
PF acknowledges support by  Funda\c{c}\~ao para a Ci\^encia e a Tecnologia (FCT) through 
the Investigador FCT contract of reference IF/01037/2013 and POPH/FSE (EC) by FEDER 
funding through the program ``Programa Operacional de Factores de Competitividade - 
COMPETE''. This work has been carried out within the frame of the National Center for Competence in Research “PlanetS” supported by the Swiss National Science Foundation (SNSF). DE acknowledges the financial support of the SNSF. NH acknowledges support from NASA Origins program through grant NNX12AQ62G and NASA ADAP program
through grant NNX13AF20G. 
\end{acknowledgements}

\bibliographystyle{aa}
\bibliography{mahlibspot}

\begin{thebibliography}{31}
\expandafter\ifx\csname natexlab\endcsname\relax\def\natexlab#1{#1}\fi

\bibitem[{{Berdyugina}(2005)}]{Berdyugina-05}
{Berdyugina}, S.~V. 2005, Living Reviews in Solar Physics, 2, 8

\bibitem[{{Boisse} {et~al.}(2009){Boisse}, {Moutou}, {Vidal-Madjar}, {Bouchy},
  {Pont}, {H{\'e}brard}, {Bonfils}, {Croll}, {Delfosse}, {Desort}, {Forveille},
  {Lagrange}, {Loeillet}, {Lovis}, {Matthews}, {Mayor}, {Pepe}, {Perrier},
  {Queloz}, {Rowe}, {Santos}, {S{\'e}gransan}, \& {Udry}}]{Boisse-09}
{Boisse}, I., {Moutou}, C., {Vidal-Madjar}, A., {et~al.} 2009, \aap, 495, 959

\bibitem[{{Bonfils} {et~al.}(2012){Bonfils}, {Gillon}, {Udry}, {Armstrong},
  {Bouchy}, {Delfosse}, {Forveille}, {Fumel}, {Jehin}, {Lendl}, {Lovis},
  {Mayor}, {McCormac}, {Neves}, {Pepe}, {Perrier}, {Pollaco}, {Queloz}, \&
  {Santos}}]{Bonfils-12}
{Bonfils}, X., {Gillon}, M., {Udry}, S., {et~al.} 2012, \aap, 546, A27

\bibitem[{{Brown} {et~al.}(2001){Brown}, {Charbonneau}, {Gilliland}, {Noyes},
  \& {Burrows}}]{Brown-01}
{Brown}, T.~M., {Charbonneau}, D., {Gilliland}, R.~L., {Noyes}, R.~W., \&
  {Burrows}, A. 2001, \apj, 552, 699

\bibitem[{{Charbonneau} {et~al.}(2002){Charbonneau}, {Brown}, {Noyes}, \&
  {Gilliland}}]{Charbonneau-02}
{Charbonneau}, D., {Brown}, T.~M., {Noyes}, R.~W., \& {Gilliland}, R.~L. 2002,
  \apj, 568, 377

\bibitem[{{Claret} \& {Bloemen}(2011)}]{Claret-11}
{Claret}, A. \& {Bloemen}, S. 2011, VizieR Online Data Catalog, 352, 99075

\bibitem[{{Crossfield} {et~al.}(2013){Crossfield}, {Barman}, {Hansen}, \&
  {Howard}}]{Crossfield-13}
{Crossfield}, I.~J.~M., {Barman}, T., {Hansen}, B.~M.~S., \& {Howard}, A.~W.
  2013, \aap, 559, A33

\bibitem[{{Czesla} {et~al.}(2009){Czesla}, {Huber}, {Wolter}, {Schr{\"o}ter},
  \& {Schmitt}}]{Czesla-09}
{Czesla}, S., {Huber}, K.~F., {Wolter}, U., {Schr{\"o}ter}, S., \& {Schmitt},
  J.~H.~M.~M. 2009, \aap, 505, 1277

\bibitem[{{Demory} {et~al.}(2013){Demory}, {Torres}, {Neves}, {Rogers},
  {Gillon}, {Horch}, {Sullivan}, {Bonfils}, {Delfosse}, {Forveille}, {Lovis},
  {Mayor}, {Santos}, {Seager}, {Smalley}, \& {Udry}}]{Demory-13}
{Demory}, B.-O., {Torres}, G., {Neves}, V., {et~al.} 2013, \apj, 768, 154

\bibitem[{{D{\'e}sert} {et~al.}(2008){D{\'e}sert}, {Vidal-Madjar}, {Lecavelier
  Des Etangs}, {Sing}, {Ehrenreich}, {H{\'e}brard}, \& {Ferlet}}]{Desert-08}
{D{\'e}sert}, J.-M., {Vidal-Madjar}, A., {Lecavelier Des Etangs}, A., {et~al.}
  2008, \aap, 492, 585

\bibitem[{{Ehrenreich} {et~al.}(2014){Ehrenreich}, Bonfils, Lovis, \&
  et~al.}]{Ehrenreich-14}
{Ehrenreich}, D., Bonfils, X., Lovis, C., \& et~al. 2014, Submitted to \aap

\bibitem[{{Ehrenreich} {et~al.}(2007){Ehrenreich}, {H{\'e}brard}, {Lecavelier
  des Etangs}, {Sing}, {D{\'e}sert}, {Bouchy}, {Ferlet}, \&
  {Vidal-Madjar}}]{Ehrenreich-07}
{Ehrenreich}, D., {H{\'e}brard}, G., {Lecavelier des Etangs}, A., {et~al.}
  2007, \apjl, 668, L179

\bibitem[{{Fukui} {et~al.}(2013){Fukui}, {Narita}, {Kurosaki}, {Ikoma},
  {Yanagisawa}, {Kuroda}, {Shimizu}, {Takahashi}, {Ohnuki}, {Onitsuka},
  {Hirano}, {Suenaga}, {Kawauchi}, {Nagayama}, {Ohta}, {Yoshida}, {Kawai}, \&
  {Izumiura}}]{Fukui-13}
{Fukui}, A., {Narita}, N., {Kurosaki}, K., {et~al.} 2013, \apj, 770, 95

\bibitem[{{Knutson} {et~al.}(2010){Knutson}, {Howard}, \&
  {Isaacson}}]{Knutson-10}
{Knutson}, H.~A., {Howard}, A.~W., \& {Isaacson}, H. 2010, \apj, 720, 1569

\bibitem[{{Knutson} {et~al.}(2011){Knutson}, {Madhusudhan}, {Cowan},
  {Christiansen}, {Agol}, {Deming}, {D{\'e}sert}, {Charbonneau}, {Henry},
  {Homeier}, {Langton}, {Laughlin}, \& {Seager}}]{Knutson-11}
{Knutson}, H.~A., {Madhusudhan}, N., {Cowan}, N.~B., {et~al.} 2011, \apj, 735,
  27

\bibitem[{{Kreidberg} {et~al.}(2014){Kreidberg}, {Bean}, {D{\'e}sert},
  {Benneke}, {Deming}, {Stevenson}, {Seager}, {Berta-Thompson}, {Seifahrt}, \&
  {Homeier}}]{Kreidberg-14}
{Kreidberg}, L., {Bean}, J.~L., {D{\'e}sert}, J.-M., {et~al.} 2014, \nat, 505,
  69

\bibitem[{{Lecavelier Des Etangs} {et~al.}(2008){Lecavelier Des Etangs},
  {Pont}, {Vidal-Madjar}, \& {Sing}}]{Lecavelier-08}
{Lecavelier Des Etangs}, A., {Pont}, F., {Vidal-Madjar}, A., \& {Sing}, D.
  2008, \aap, 481, L83

\bibitem[{{Meunier} {et~al.}(2010){Meunier}, {Desort}, \&
  {Lagrange}}]{Meunier-10}
{Meunier}, N., {Desort}, M., \& {Lagrange}, A.-M. 2010, \aap, 512, A39

\bibitem[{{Nascimbeni} {et~al.}(2013){Nascimbeni}, {Piotto}, {Pagano},
  {Scandariato}, {Sani}, \& {Fumana}}]{Nascimbeni-13}
{Nascimbeni}, V., {Piotto}, G., {Pagano}, I., {et~al.} 2013, \aap, 559, A32

\bibitem[{{Oshagh} {et~al.}(2013{\natexlab{b}}){Oshagh}, {Boisse}, {Bou{\'e}},
  {Montalto}, {Santos}, {Bonfils}, \& {Haghighipour}}]{Oshagh-13b}
{Oshagh}, M., {Boisse}, I., {Bou{\'e}}, G., {et~al.} 2013{\natexlab{b}}, \aap,
  549, A35

\bibitem[{{Oshagh} {et~al.}(2013{\natexlab{a}}){Oshagh}, {Santos}, {Boisse},
  {Bou{\'e}}, {Montalto}, {Dumusque}, \& {Haghighipour}}]{Oshagh-13a}
{Oshagh}, M., {Santos}, N.~C., {Boisse}, I., {et~al.} 2013{\natexlab{a}}, \aap,
  556, A19

\bibitem[{{Pont} {et~al.}(2007){Pont}, {Gilliland}, {Moutou}, {Charbonneau},
  {Bouchy}, {Brown}, {Mayor}, {Queloz}, {Santos}, \& {Udry}}]{Pont-07}
{Pont}, F., {Gilliland}, R.~L., {Moutou}, C., {et~al.} 2007, \aap, 476, 1347

\bibitem[{{Pont} {et~al.}(2008){Pont}, {Knutson}, {Gilliland}, {Moutou}, \&
  {Charbonneau}}]{Pont-08}
{Pont}, F., {Knutson}, H., {Gilliland}, R.~L., {Moutou}, C., \& {Charbonneau},
  D. 2008, \mnras, 385, 109

\bibitem[{{Pont} {et~al.}(2013){Pont}, {Sing}, {Gibson}, {Aigrain}, {Henry}, \&
  {Husnoo}}]{Pont-13}
{Pont}, F., {Sing}, D.~K., {Gibson}, N.~P., {et~al.} 2013, \mnras, 432, 2917

\bibitem[{{Poppenhaeger} {et~al.}(2013){Poppenhaeger}, {Schmitt}, \&
  {Wolk}}]{Poppenhaeger-13}
{Poppenhaeger}, K., {Schmitt}, J.~H.~M.~M., \& {Wolk}, S.~J. 2013, \apj, 773,
  62

\bibitem[{{Seager} \& {Sasselov}(2000)}]{Seager-00}
{Seager}, S. \& {Sasselov}, D.~D. 2000, \apj, 537, 916

\bibitem[{{Sing} {et~al.}(2009){Sing}, {D{\'e}sert}, {Lecavelier Des Etangs},
  {Ballester}, {Vidal-Madjar}, {Parmentier}, {Hebrard}, \& {Henry}}]{Sing-09}
{Sing}, D.~K., {D{\'e}sert}, J.-M., {Lecavelier Des Etangs}, A., {et~al.} 2009,
  \aap, 505, 891

\bibitem[{{Sing} {et~al.}(2011){Sing}, {Pont}, {Aigrain}, {Charbonneau},
  {D{\'e}sert}, {Gibson}, {Gilliland}, {Hayek}, {Henry}, {Knutson}, {Lecavelier
  Des Etangs}, {Mazeh}, \& {Shporer}}]{Sing-11}
{Sing}, D.~K., {Pont}, F., {Aigrain}, S., {et~al.} 2011, \mnras, 416, 1443

\bibitem[{{Solanki}(2003)}]{Solanki-03}
{Solanki}, S.~K. 2003, \aapr, 11, 153

\bibitem[{{Unruh} {et~al.}(1999){Unruh}, {Solanki}, \& {Fligge}}]{Unruh-99}
{Unruh}, Y.~C., {Solanki}, S.~K., \& {Fligge}, M. 1999, \aap, 345, 635

\bibitem[{{Worden} {et~al.}(1998){Worden}, {White}, \& {Woods}}]{Worden-98}
{Worden}, J.~R., {White}, O.~R., \& {Woods}, T.~N. 1998, \apj, 496, 998

\end{thebibliography}

\end{document}